\documentclass[aps,prb,amsmath,amssymb,twocolumn,superscriptaddress,floatfix]{revtex4-1}
\usepackage{graphicx}
\usepackage{hyperref}
\hypersetup{pdftitle={BiPd ARPES},pdfauthor={Sasha Yaresko},pdfsubject={Superconductivity},pdfdisplaydoctitle}

\usepackage{color}

\begin{document}

\title{Correct Brillouin zone and electronic structure of {BiPd}}

\author{Alexander Yaresko}
\email{a.yaresko@fkf.mpg.de}

\author{Andreas P.\ Schnyder}
\email{a.schnyder@fkf.mpg.de}

\author{Hadj M.\ Benia}
\affiliation{Max-Planck-Institut f\"ur Festk\"orperforschung, Heisenbergstr.\ 1, 70569 Stuttgart, Germany}

\author{Chi-Ming Yim}
\affiliation{SUPA, School of Physics and Astronomy, University of St.\ Andrews, North Haugh, St.\ Andrews, Fife KY16 9SS, United Kingdom}

\author{Giorgio Levy}
\author{Andrea Damascelli}
\affiliation{Department of Physics and Astronomy, University of British Columbia, Vancouver, British Columbia V6T 1Z1, Canada}

\author{Christian R.\ Ast}
\affiliation{Max-Planck-Institut f\"ur Festk\"orperforschung, Heisenbergstr.\ 1, 70569 Stuttgart, Germany}

\author{Darren C.\ Peets}
\email{dpeets@fudan.edu.cn}
\affiliation{Department of Physics and Advanced Materials Laboratory, Fudan University, 2205 SongHu Road, Shanghai 200438, China}

\author{Peter Wahl}
\affiliation{SUPA, School of Physics and Astronomy, University of St.\ Andrews, North Haugh, St.\ Andrews, Fife KY16 9SS, United Kingdom}

\begin{abstract}
A promising route to the realization of Majorana fermions is in non-centrosymmetric superconductors, in which spin-orbit-coupling lifts the spin degeneracy of both bulk and surface bands. A detailed assessment of the electronic structure is critical to evaluate their suitability for this through establishing the topological properties of the electronic structure. This requires correct identification of the time-reversal-invariant momenta. One such material is BiPd, a recently rediscovered non-centrosymmetric superconductor which can be grown in large, high-quality single crystals and has been studied by several groups using angular resolved photoemission to establish its surface electronic structure.  Many of the published electronic structure studies on this material are based on a reciprocal unit cell which is not the actual Brillouin zone of the material. We show here the consequences of this for the electronic structures and show how the inferred topological nature of the material is affected.  
\end{abstract}

\maketitle

\section{Introduction}

In topologically-nontrivial materials without inversion symmetry, spin-orbit coupling splits the spin degeneracy of the bulk bands leading to a comlex \mbox{(pseudo-)spin} texture of the electronic wave functions. 
This, in turn,  gives rise to non-trivial topological properties of the band structure, such as Dirac or Weyl band-crossing points~\cite{chiu_review_RMP,armitage_mele_vishwanath_review}, protected surface states~\cite{hasan:rmp,qi:rmp}, and, in the case of superconductors, 
Majorana quasiparticles which are their own antiparticle~\cite{elliott_franz_RMP_majorana_15,schnyder_brydon_review_JPCM_15}. 
A major thrust of current research focuses on the interplay between topologically nontrivial surface states and superconductivity.
This provides opportunities to create Majorana fermions~\cite{fu_kane_PRL_08,he_zhang_chiral_majorana_science_17}, as well as topologically-nontrivial superconducting states~\cite{shin_ding_arXiv_iron_based_SC_topo}, and the possibilities of topological transitions driven by pressure or magnetic field.
 Few systems are known which exhibit both topologically-nontrivial band structure and superconductivity~\cite{shin_ding_arXiv_iron_based_SC_topo,bian_hasan_PbTaSe2_nat_commun_16,Mondal2012}. BiPd in its low temperature phase, $\alpha$-BiPd, is one such material~\cite{Ionov1989,Bhatt1979}.  
This material is also a noncentrosymmetric superconductor\cite{Alekseevskii1952,Kheiker1953}, in which the superconducting condensate is expected to have mixed singlet and triplet character\cite{Bauerbook,Fujimoto2007}. That such topologically-nontrivial superconducting pairing might be realized in BiPd has led to a recent renaissance of this material\cite{joshi2011b,Mondal2012}. In particular, BiPd has been shown to exhibit surface states with Dirac-like dispersion in directional bandgaps, despite having a large number of bands which cross the Fermi level\cite{Sun2015,Neupane2016,Benia2016,Setti2016}.  Because of the low symmetry of the material, the surface states of BiPd are different on the [010] and [0\=10] surfaces\cite{Benia2016}.  

Recently, detailed photoemission characterizations of the surface electronic structure of $\alpha$-BiPd have been published in several independent studies.\cite{Neupane2016,Benia2016,Setti2016,Lohani2017} While the data broadly agree, the interpretation of the band structure does not. Neupane {\it et al.}~\cite{Neupane2016} and Setti {\it et al.}\cite{Setti2016} place all surface states at the $\Gamma$ point, consistent with their band structure calculations, whereas Benia {\it et al.}~\cite{Benia2016} locate those below the Fermi level at the $S$ point and those above the Fermi level at the $\Gamma$ point, consistent with their own band structure calculations~\cite{Benia2016,Sun2015}. Here, we set out to address these differing interpretations, and clarify how they are related --- tracing them back to the unit cell used for electronic structure calculations. 

Clearly, these assignments cannot both be correct. A key difference is in the underlying unit cell used for calculations and interpretation of the data: Benia {\it et al}.\ use the primitive cell as reported by Ionov {\it et al}.\cite{Ionov1989}, whereas Neupane {\it et al}. and Setti {\it et al}. use the larger, non-primitive unit cell reported by Bhatt {\it et al}.\cite{Bhatt1979}. This latter unit cell is also used by Lohani {\it et al}., although they concentrate on bulk states\cite{Lohani2017}.  This doubled cell has a reciprocal cell half the size of its Brillouin Zone.  Here we show that the two unit cells yield very similar band structures, but only if the correct Brillouin zone is used for the unconventional base-centered $B2_1$ structure of Bhatt {\it et al}. Using the correct Brillouin zone places the surface bands in the occupied states at the $S$ point, whereas using a larger unit cell folds these bands to the zone center ($\Gamma$ point). Most importantly, in the Brillouin zone a directional band gap opens due to spin-orbit coupling at the $\Gamma$ point in the unoccupied states which is filled with bands folded in from the zone face if the incorrect unit cell is used.

\section{Methods}

\begin{figure*}[ht]
  \includegraphics[width=\textwidth]{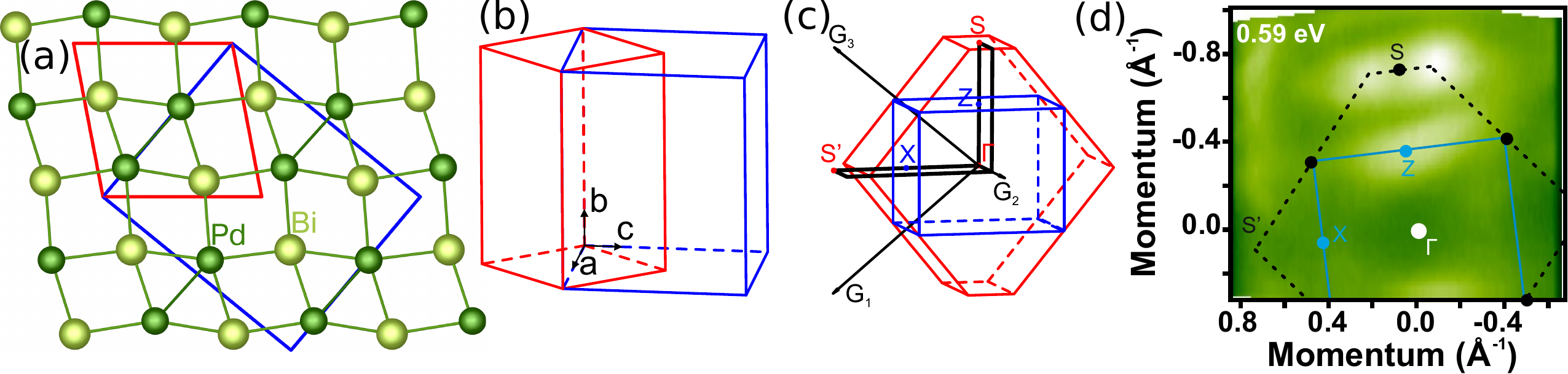}
  \caption{\label{fig1a} The monoclinic $P2_1$ and the pseudo-orthorhombic $B2_1$ unit cells. Red lines show the (a,b) $P2_1$ unit cell in real space and (c) its reciprocal unit cell, which is also the Brillouin zone of both the $P2_1$ and $B2_1$ cells.  The blue lines indicate the doubled $B2_1$ unit cell and its smaller reciprocal unit cell as reported in Ref.\ \onlinecite{Bhatt1979}.  This reciprocal cell is half the size of the Brillouin zone because of the doubled, non-primitive unit cell in real space.  The $z=0$ plane in $P2_1$ is shown in (a).  (d) Comparison of the Brillouin zone and the $B2_1$ reciprocal cell with ARPES data\cite{Benia2016}.}
\end{figure*}

We employed fully-relativistic linear-muffin-tin-orbital calculations using the crystal structures by Bhatt {\it et al}.\cite{Bhatt1979} and Ionov {\it et al}.\cite{Ionov1989}. Details of the calculations have been reported in Refs.\ \onlinecite{Benia2016} and \onlinecite{Sun2015}. The crystal used for angle-resolved photoemission spectroscopy (ARPES) measurements was grown by a modified Bridgman-Stockbarger method as reported in detail elsewhere\cite{Peets2014}.  ARPES was performed on a freshly cleaved surface using a \mbox{Helium-I} source ($\nu=21.2$\,eV) with a hemispherical SPECS HSA3500 electron analyzer. Throughout this paper we refer to $\alpha$-BiPd simply as BiPd since the $\beta$ phase has not been stabilized at or below room temperature. 

\section{Results and Discussion}

A crystal structure's Brillouin zone is a uniquely-defined primitive cell in reciprocal space.  Its reciprocal cell is a region in reciprocal space corresponding to the unit cell in real space. These are often, but not always, the same.  If the real-space unit cell is a primitive cell, its reciprocal cell will correspond to the Brillouin zone. However, for convenience a unit cell is sometimes chosen to have alternative orientations, settings, or sizes, which can affect the reciprocal cell but {\slshape not} the Brillouin zone. This should be intuitively clear: An arbitrary choice of labelling cannot change the physical properties of a material, for instance by altering the Fermi surface, or by converting an indirect-bandgap semiconductor into a direct-bandgap semiconductor through sufficient band folding.  BiPd is an example of a system in which one must be aware of this distinction, since some authors chose a doubled pseudo-orthorhombic supercell to aid comparison to related materials --- see Fig.~\ref{fig1a}(a,b). This cell is convenient, but larger and non-primitive, and its reciprocal cell is half the size of the Brillouin Zone.  In Figure~\ref{fig1a}(c) we show the reciprocal unit cell of BiPd for the larger pseudo-orthorhombic unit cell of Bhatt {\it et al}.\ and for the primitive cell of Ionov {\it et al}.  The latter reciprocal cell is the correct Brillouin zone for both crystal structures. Experimentally, this can be verified from the size of the observed Brillouin zone: the reciprocal cell of the doubled $B2_1$ unit cell is half the size of the Brillouin zone, with the X and Z points at 0.5\,\AA$^{-1}$ and 0.41\,\AA$^{-1}$, respectively. Comparison with ARPES constant-energy maps [Figure~\ref{fig1a}(d)] confirms that features repeat on the scale of the Brillouin zone and not on the scale of the smaller reciprocal cell corresponding to the unit cell used by Bhatt. It is worth noting that a surface reconstruction would lead to the observation of a smaller Brillouin zone than what has been reported experimentally by any group. 

\begin{figure}[ht]
  \includegraphics[width=\columnwidth]{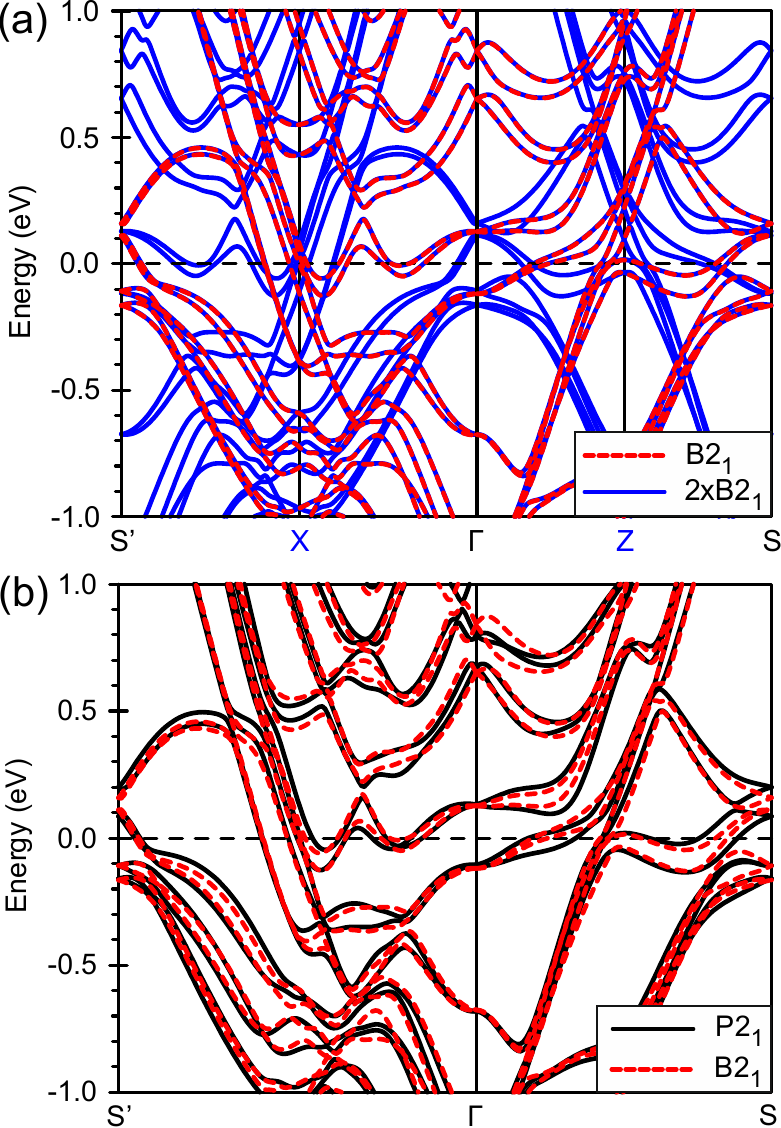}
  \caption{\label{fig1b} Electronic structure in $P2_1$ and $B2_1$. (a) Band structure of BiPd calculated for $B2_1$ in its Brillouin zone (red dashed lines) and taking the reciprocal cell of the doubled unit cell as the Brillouin zone (blue solid line). For the latter, $X$ and $Z$ are at the zone boundary. (b) The band structure in the same energy range as in (a), obtained in the Brillouin zone for both crystal structures, {\it i.e}.\ $B2_1$ (red dashed lines) and $P2_1$ (black solid lines).}
\end{figure}

Figure~\ref{fig1b}(a) shows the band structure calculated for the base-centered $B2_1$ unit cell defined by Bhatt {\it et al}.\ (red dashed lines) using its Brillouin zone, and for a doubled primitive pseudo-orthorhombic unit cell having the same lattice parameters and monoclinic angle but a Brillouin zone half as large. For this calculation, in other words, the larger blue unit cell in Fig.~\ref{fig1a}(a,b) is treated as primitive and the smaller blue reciprocal cell in Fig.~\ref{fig1a}(c) is thus treated as a Brillouin zone.  With twice as many basis states in its primitive real space cell, the electronic structure for the doubled pseudo-orthorhombic cell shows twice as many bands as the base-centered cell in reciprocal space. Half of these bands exhibit exactly the same dispersion along X--$\Gamma$--Z as for the base-centered one, with additional bands being folded to $\Gamma$--X ($\Gamma$--Z) from S$'$--X (S--Z). This is to be compared with Supplementary Fig.~5 (right panel) of Neupane's paper\cite{Neupane2016}, showing excellent agreement. It is worth noting that Setti {\it et al}.\ use an orthorhombic rather than a monoclinic unit cell, which is correct only for the high temperature phase $\beta$-BiPd. In Figure~\ref{fig1b}(b) we plot for comparison the band structure for the unit cells of Bhatt \textit{et al.} and Ionov \textit{et al.} ($P2_1$), used by Benia {\it et al}.\cite{Benia2016}, on the same energy and momentum scale as in Fig.~\ref{fig1b}(a), showing near-perfect agreement between bands for the two different crystal structures. It can be seen that using the correct Brillouin zone leads to a much smaller number of bands at the $\Gamma$ point and preserves the directional band gap near the $\Gamma$ point above the Fermi level.

\begin{figure}[htb]
  \includegraphics[width=\columnwidth]{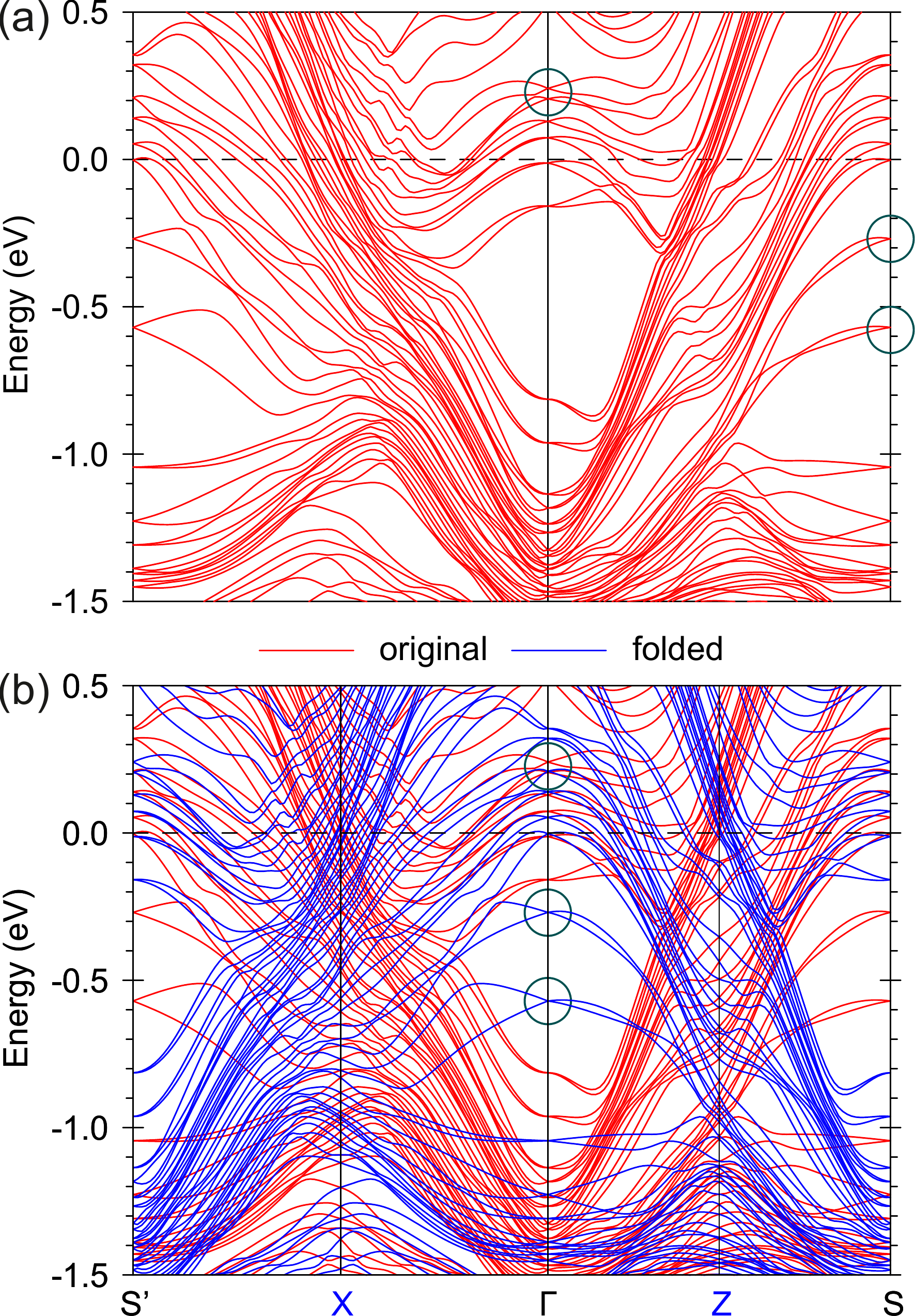}
  \caption{\label{fig2}Effect of zone folding on the surface band structure.  (a) Band structure obtained for a slab and shown in the surface Brillouin zone of the $P2_1$ crystal structure\cite{Sun2015}. Circles mark the four surface states obtained in the slab calculation, two in the occupied states at S and two in the unoccupied states at $\Gamma$.  (b) Band structure shown in the reduced reciprocal unit cell for the $B2_1$ structure with the zone boundaries at X and Z, which is not a Brillouin zone. Shown in red are the bands from the unfolded band structure, the ones shown in blue appear in addition due to back-folding at X and Z. The Dirac-like surface states are identified with circles in (a) at their position in the Brillouin zone and in (b) where they appear after folding. The surface states above the Fermi energy near $\Gamma$, which were originally in a directional bandgap opened by spin-orbit coupling, are enveloped in folded bands, whereas the ones which were at $S$ in the Brillouin zone (a) now appear at $\Gamma$.}
\end{figure}

Figure~\ref{fig2} shows the results of a slab calculation performed in the $P2_1$ Brillouin zone, with the surface Dirac-like states identified by circles.  If the original bands, shown in red, are folded to match the reciprocal cell of Bhatt {\it et al}., the blue bands result.  This folds all the surface states to the $\Gamma$ point.  Crucially, the surface states located above the Fermi energy are no longer in a directional bandgap opened by spin-orbit coupling, and overlap with folded bands.  This would change their interpretation.  It would also most likely render them impossible to observe.   Meanwhile, folded bands cross the surface bands on top of the van Hove singularity detected previously in STS\cite{Sun2015}.  The interaction with the folded bulk bands would eliminate this van Hove singularity, and the fact that it is, in fact, observed experimentally demonstrates that the folded picture cannot be correct.  This is one example of the direct experimental consequences of band folding mentioned above --- there is no freedom to choose an arbitrary Brillouin zone.

The correct assignment of the Brillouin zone is important for two reasons: First, as shown in Fig.~\ref{fig2}, the surface states above the Fermi level become merely surface resonances if calculated in the larger unit cell, due to bands being folded into its directional bandgap at the $\Gamma$ point.  Second and most crucially, a reliable assessment of the topological nature of the material is not possible without a correct assignment of the time-reversal-invariant momenta. This is true in particular for the directional bandgap at the $\Gamma$ point, which is opened by spin-orbit coupling and is central to the topological nature of BiPd.  If this directional bandgap no longer separates states split by spin-orbit coupling, the surface states are no longer topological in nature.  Doubling the Brillouin zone may constitute a topological transition of the lattice.

The confusion over the correct Brillouin zone may ultimately stem from early reports which identified the structure as $Ccm2_1$ (Tab.\ \ref{tab:Kheiker})\cite{Kheiker1953}, later refined to $C_2^2$ ($B2_1$) by Bhatt \textit{et al.} (Tab.~\ref{tab:Bhatt}), a choice which facilitates comparison to the crystal structure of thallium(I) iodide (TlI)\cite{Bhatt1979,Kheiker1953} --- this, however, is not a primitive cell. More-recent characterization by x-ray diffraction has identified the primitive unit cell as being of $P2_1$ symmetry (Tab.\ \ref{tab:Ionov293})~\cite{Ionov1989}, although this report is not available online.  To aid future work on BiPd, the results of these structure refinements are reproduced in Appendix \ref{App}.

\section{Conclusion}

In summary, a correct treatment using the Brillouin zone rather than the reciprocal unit cell places the topologically-protected surface states below the Fermi level of the [010] and [0\=10] surfaces near the S point, whereas those above the Fermi level reside in a directional band gap opened by spin-orbit coupling at the zone center. The correct location of the surface states is crucial for understanding the physics in this material.

\begin{acknowledgments}
Support from the MPG-UBC Centre, the Canadian Institute for Advanced Research, the Engineering and Physical Sciences Research Council (EP/I031014/1), and the National Natural Science Foundation of China (Project No.~11650110428) is gratefully acknowledged.  This work was supported by the DFG within projects STA315/8-1 and BE5190/1-1.  

\end{acknowledgments}

\appendix

\section{Reported Crystal Structures\label{App}}

Previously-reported crystal structures are reproduced here for reference.  Most are not available online or in English at present, and an inability to access this information has likely impeded band structure calculations and the interpretation of surface spectroscopy data.

\begin{table}[h!]
  \caption{\label{tab:Kheiker}Crystal structure of BiPd as refined by Kheiker\cite{Kheiker1953}, with lattice parameters $a$\,=\,7.203(1)\,\AA, $b$\,=\,8.707(2)\,\AA, and $c$\,=\,10.662(1)\,\AA, in the space group $Ccm2_1$, an alternate setting of $Cmc2_1$ (No.~36).}
  \begin{tabular}{lclll}\hline
    Site & Wyckoff & \multicolumn{1}{c}{$x/a$} & \multicolumn{1}{c}{$y/b$} & \multicolumn{1}{c}{$z/c$}\\ \hline\hline
    Bi1 & 4a & 0.108 & 0. & 0.\\
    Bi2 & 8b & 0.125 & 0.274 & 0.722\\
    Bi3 & 4a & 0.650 & 0. & 0. \\
    Pd1 & 4a & 0.108 & 0. & 0.275\\
    Pd2 & 8b & 0.125 & 0.274 & 0.447\\
    Pd3 & 4a & 0.650 & 0. & 0.275\\ \hline
  \end{tabular}
\bigskip
%
%

  \caption{\label{tab:Bhatt}Crystal structure of BiPd as refined by Bhatt\cite{Bhatt1979}, with lattice parameters $a$\,=\,7.200(1)\,\AA, $b$\,=\,10.660(4)\,\AA, $c$\,=\,8.708(2)\,\AA, and $\beta$\,=\,89.70(3)$^\circ$, in $B2_1$, an unconventional setting of $P2_1$ (No.~4).  We have not reproduced the anisotropic thermal parameters here.}
  \begin{tabular}{lclll}\hline
    Site & Wyckoff & \multicolumn{1}{c}{$x/a$} & \multicolumn{1}{c}{$y/b$} & \multicolumn{1}{c}{$z/c$}\\ \hline\hline
    Bi1 & 2a & 0.1193(6) & 0. & 0.9874(6)\\
    Bi2 & 2a & 0.6394(6) & 0.9937(8) & 0.0118(6)\\
    Bi3 & 2a & 0.1232(6) & 0.7123(5) & 0.2752(4) \\
    Bi4 & 2a & 0.6185(6) & 0.7139(7) & 0.2256(4) \\
    Pd1 & 2a & 0.086(1) & 0.271(1) & 0.001(1)\\
    Pd2 & 2a & 0.676(1) & 0.265(1) & 0.001(1)\\
    Pd3 & 2a & 0.146(2) & 0.437(1) & 0.260(1)\\
    Pd4 & 2a & 0.594(2) & 0.441(1) & 0.242(1)\\ \hline
  \end{tabular}
\bigskip
  \caption{\label{tab:Bhatt_p1}Crystal structure of BiPd as refined by
  Bhatt\cite{Bhatt1979} recalculated to conventional setting of $P2_1$ (No.~4), with lattice parameters $a$\,=\,5.635\,\AA, $b$\,=\,5.664\,\AA, $c$\,=\,10.660\,\AA, and $\gamma$\,=\,100.83(3)$^\circ$.}
  \begin{tabular}{lclll}\hline
    Site & Wyckoff & \multicolumn{1}{c}{$x/a$} & \multicolumn{1}{c}{$y/b$} & \multicolumn{1}{c}{$z/c$}\\ \hline\hline
    Bi1 & 2a & 0.1319 & 0.1067 & 0.0 \\
    Bi2 & 2a & 0.6276 & 0.6512 & 0.9937\\
    Bi3 & 2a & -0.1520 & 0.3984 & 0.7123 \\
    Bi4 & 2a & 0.3929 & 0.8441 & 0.7139 \\
    Pd1 & 2a & 0.085 & 0.087 & 0.271\\
    Pd2 & 2a & 0.675 & 0.677 & 0.265\\
    Pd3 & 2a & -0.114 & 0.406 & 0.437\\
    Pd4 & 2a & 0.352 & 0.836 & 0.441 \\ \hline
  \end{tabular}
\end{table}%
\begin{table}[h!]
  \caption{\label{tab:Ionov293}Crystal structure of $\alpha$-BiPd at 293\,K as refined by Ionov\cite{Ionov1989}, with lattice parameters $a$\,=\,5.635(2)\,\AA, $b$\,=\,5.661(2)\,\AA, $c$\,=\,10.651(5)\,\AA, and $\gamma$\,=\,100.85(3)$^\circ$, in $P2_1$ (No.~4).}
  \begin{tabular}{lcllr@{.}lr@{.}l}\hline
    Site & Wyckoff & \multicolumn{1}{c}{$x/a$} & \multicolumn{1}{c}{$y/b$} & \multicolumn{2}{c}{$z/c$} & \multicolumn{2}{c}{$B$ (\AA$^2$)}\\ \hline\hline
    Bi1 & 2a & 0.6565(6) & 0.1055(6) & 0& & 0&6\\
    Bi2 & 2a & 0.1160(4) & 0.6545(4) & 0&0050(3) & 0&4\\
    Bi3 & 2a & 0.6360(4) & 0.6014(5) & -0&2116(3) & 0&1\\
    Bi4 & 2a & 0.1129(6) & 0.1529(5) & -0&2163(4) & 0&7\\
    Pd1 & 2a & 0.406(1) & 0.908(2) & 0&235(1) & 1&4\\
    Pd2 & 2a & 0.858(1) & 0.346(1) & 0&241(1) & 0&3\\
    Pd3 & 2a & 0.598(1) & 0.589(1) & 0&068(1) & 0&2\\
    Pd4 & 2a & 0.178(1) & 0.191(1) & 0&064(1) & 0&8\\ \hline
  \end{tabular}
\bigskip
  \caption{\label{tab:Ionov493}Crystal structure of $\beta$-BiPd at 493\,K as refined by Ionov\cite{Ionov1989}, with lattice parameters $a$\,=\,5.674(2)\,\AA, $b$\,=\,5.691(2)\,\AA, $c$\,=\,10.596(4)\,\AA, and $\gamma$\,=\,101.4(1)$^\circ$, again in $P2_1$ (No.~4).}
  \begin{tabular}{lcllr@{.}lr@{.}l}\hline
    Site & Wyckoff & \multicolumn{1}{c}{$x/a$} & \multicolumn{1}{c}{$y/b$} & \multicolumn{2}{c}{$z/c$} & \multicolumn{2}{c}{$B$ (\AA$^2$)}\\ \hline\hline
    Bi1 & 2a & 0.667(1) & 0.142(2) & 0& & 1&1\\
    Bi2 & 2a & 0.143(4) & 0.648(3) & 0&034(1) & 1&6\\
    Bi3 & 2a & 0.623(6) & 0.589(5) & -0&166(2) & 4&7\\
    Bi4 & 2a & 0.121(3) & 0.109(2) & -0&213(1) & 0&6\\
    Pd1 & 2a & 0.457(6) & 0.950(5) & 0&253(3) & 1&4\\
    Pd2 & 2a & 0.857(2) & 0.402(2) & 0&222(2) & 1&6\\
    Pd3 & 2a & 0.677(2) & 0.673(2) & 0&086(2) & 1&7\\
    Pd4 & 2a & 0.107(2) & 0.155(2) & 0&039(1) & 2&3\\ \hline
  \end{tabular}
\end{table}
\bibliography{BiPd_comment,BiPd}

\end{document}